\documentclass[bimj,fleqn]{w-art}

\usepackage{listings}

\usepackage[utf8]{inputenc}
\usepackage[T1]{fontenc}
\usepackage{lmodern}
\usepackage{textcomp}
\usepackage{color}
\usepackage{enumerate}
\usepackage{graphicx}
\usepackage{grffile}
\usepackage{wrapfig}
\usepackage{rotating}
\usepackage{longtable}
\usepackage{multirow}
\usepackage{multicol}
\usepackage{changes}
\usepackage{pdflscape}
\usepackage{geometry}
\usepackage[normalem]{ulem}
\usepackage{amssymb}
\usepackage{amsmath}
\usepackage{amsfonts}
\usepackage{dsfont}
\usepackage{array}
\usepackage{ifthen}
\usepackage{hyperref}
\usepackage{natbib}
\title[Ozenne]{On the estimation of average treatment effects with right-censored time to event outcome and competing risks}
\author[Ozenne]{Brice Maxime Hugues Ozenne\inst{1,2}}
\author[Scheike]{Thomas Harder Scheike\inst{1}\footnote{Corresponding author: {\sf{e-mail: ts@biostat.ku.dk}}, Phone: +00-999-999-999, Fax: +00-999-999-999}}
\author[St\ae rk]{Laila St\ae rk\inst{4}}
\author[Gerds]{Thomas Alexander Gerds\inst{1,5}}
\address[\inst{1}]{Department of Biostatistics, University of Copenhagen, Copenhagen, Denmark}
\address[\inst{2}]{Neurobiology Research Unit, University Hospital of Copenhagen, Rigshospitalet, Copenhagen, Denmark}
\address[\inst{3}]{University Hospital of Copenhagen, Rigshospitalet, Copenhagen, Denmark}
\address[\inst{4}]{Department of Cardiology, Copenhagen University Hospital Herlev and Gentofte, Hellerup, Denmark}
\address[\inst{5}]{Danish Heart Foundation, Copenhagen, Denmark}
\usepackage[]{graphicx}
\usepackage{array}
\usepackage{lscape}
\usepackage{xspace}
\usepackage{endfloat}
\newtheorem{theorem}{Theorem}
\RequirePackage{xifthen}
\newcommand\T{T} 
\newcommand\C{C} 
\newcommand\Tobs{\tilde{T}} 
\newcommand\Cause{\ensuremath{\Delta}} 
\newcommand\Etype{\ensuremath{\tilde\Delta}} 
\newcommand\Treat{A} 
\newcommand\treat{a} 
\newcommand\X{\ensuremath{W}} 
\newcommand\x{\ensuremath{w}} 
\newcommand\XSupport{\ensuremath{\mathcal W}} 
\newcommand\Y{Y} 
\newcommand\sample{O} 
\newcommand\ATE{\Psi} 
\newcommand\Esp{\ensuremath{\mathrm E}}
\newcommand\Prob{\ensuremath{\mathrm P}}
\newcommand\dt{\ensuremath{\mathrm {d}t}}

\newcommand\Ind[1][]{
\mathrm{1}{\lbrace #1 \rbrace}
}
\newcommand\Ymodel{F_1}
\newcommand\hatYmodel{\ensuremath{\hat F_{1n}}}
\newcommand\Ftwo{F_2}
\newcommand\AIPTW{\ensuremath{\operatorname{AIPTW}}\xspace}
\newcommand\IPTW{\ensuremath{\operatorname{IPTW}}\xspace}
\newcommand\AIPCW{\ensuremath{\operatorname{AIPCW}}\xspace}
\newcommand\IPCW{\ensuremath{\operatorname{IPCW}}\xspace}
\newcommand\Tmodel{\pi}
\newcommand\hatTmodel{\ensuremath{\hat\pi_n}}
\newcommand\Cmodel{G}
\newcommand\hatCmodel{\ensuremath{\hat G_n}}
\newcommand\Smodel{\ensuremath{S}}
\newcommand\hatSmodel{\ensuremath{\hat S_n}}

\newcommand\Score{m}
\newcommand\IF{\ensuremath{\mathrm{IF}}}
\newcommand\IFY{\ensuremath{\mathrm{IF_{\Ymodel^*}}}}
\newcommand\IFS{\ensuremath{\mathrm{IF_{\Smodel^*}}}}
\newcommand\IFL[1][]{\ensuremath{\mathrm{IF}_{\Lambda_{#1}^*}}}
\newcommand\IFC{\ensuremath{\mathrm{IF_{\Cmodel^*}}}}
\newcommand\IFT{\ensuremath{\mathrm{IF_{\Tmodel^*}}}}

\date{\today}
\title{}
\hypersetup{
 colorlinks=true,
 citecolor=[rgb]{0,0.5,0},
 urlcolor=[rgb]{0,0,0.5},
 linkcolor=[rgb]{0,0,0.5},
 pdfauthor={Brice Ozenne},
 pdftitle={},
 pdfkeywords={},
 pdfsubject={},
 pdfcreator={Emacs 25.2.1 (Org mode 9.0.4)},
 pdflang={English}
 }
\begin{document}

\begin{abstract}
We are interested in the estimation of average treatment effects based on right-censored data of an observational study.
We focus on causal inference of differences between t-year absolute event risks
in a situation  with competing risks. 
We derive doubly robust estimation equations and implement estimators for the nuisance parameters
based on working regression models for the outcome, the censoring and the treatment distribution conditional on 
auxiliary baseline covariates.
We use the functional delta method to show that our estimators are regular asymptotically linear 
estimators and estimate their variances based on estimates of their influence functions.
In empirical studies we assess the robustness of the estimators and 
the coverage of confidence intervals.
The methods are further illustrated using data from a 
Danish registry study. 
\end{abstract}
\title[Average treatment effects with right-censored data and competing risks]{On the estimation of average treatment effects with right-censored time to event outcome and competing risks}
\maketitle

\section{Introduction}
\label{sec:org7a317d6}

Average treatment effects (ATE) are important parameters in
epidemiology \citep{robins1986new,hernan2006instruments}. In
observational studies, these parameters are interpreted in a suitable
framework for causal inference \citep{hernanrobinsbook,Perl2000} as what
one would have observed had the treatment been randomized. Estimators
of average treatment effects include outcome regression model based
estimators which standardize the expected outcome to a given
distribution of the confounders (G-formula), inverse probability of
treatment weighted (IPTW) estimators which rely on a model for the
propensity of treatment, and doubly robust estimators which combine
the two types of estimators with the aim to reduce bias
\citep{laan:robins:book03,kang2007demystifying,glynn2010introduction}.

In this article, we are motivated by applications in
pharmacoepidemiology where the aim is to evaluate differences between
alternative drug treatments based on large scale registry data
\citep{hernan2016using}. We are particularly interested in applications
where the outcome is a right censored time to event and death without
the outcome is a competing risk. We restrict our discussion to the
simple setting where a treatment decision is made only once at
baseline and all patients are supposed to stay on treatment for a
pre-specified amount of time, say until \(\tau\)-years after
initiation of the treatment. We then consider treatment differences
between the absolute \(\tau\)-year risks of an event of interest as
the main target parameter and aim to achieve an interpretation in the
counterfactual world where the treatment is randomized.

Compared to other approaches for competing risk data
\citep{andersen2017causal,bekaert2010adjusting,moodie2014marginal}, the
major difference of our approach is that we use working Cox regression
models for the cause-specific hazard rates to estimate the absolute
risk of the event
\citep{benichou1990estimates,ozenne2017riskregression}. Furthermore, we
allow the censoring distribution to depend on baseline covariates via
a separate Cox regression model and work with a logistic regression
model for the propensity of treatment.  We study the robustness of our
estimator to a possible misspecification of any of these working
models. Our work relates and extends recent developments in survival
analysis: \cite{wang2016double} proposed a doubly robust estimator for
right-censored survival data when using parametric working regression
models for the outcome distribution and the treatment distribution,
and a non-parametric model (Kaplan Meier) for the censoring
distribution. Using the semiparametric theory
\citep{bickel,laan:robins:book03,tsiatis2006semiparametric}, we derive
an augmentation term which makes our estimator robust against
misspecification of the censoring model. The augmentation term
resembles the one in the survival case \citep{zhao2014doubly}. We also
derive the influence function of our estimator and show that it can be
greatly simplified when all working models are correctly specified.

 This paper is structured as follows: Section \ref{CRsetting} formally
introduces the competing risk setting, the parameter of interest, and
the statistical models. Section \ref{estimatorATE} presents the G-formula,
IPTW, and doubly robust estimators in a competing risk setting. We
derive in section \ref{asymptotic} the asymptotic properties of our three
estimators: consistency, asymptotic normality, and their influence
function. Robustness of the estimators to model misspecification and
coverage of confidence intervals based on the asymptotic distribution
of the estimators is assessed in section \ref{simulation} using simulation
studies. Finally, in section \ref{application}, we apply our estimators to
compare two anticoagulation treatments regarding their impact on the
risk of bleeding (adverse endpoint) in patients with atrial
fibrillation. The data used for this illustration are a subset of the
data of \citet{staerk2018}, where we applied Cox regression for the event
hazard and the hazard of death without event in order to estimate
average differences in \(\tau\)-year risk of stroke and bleeding between
alternative drugs for anticoagulation therapy.

\section{Competing risk setting}
\label{CRsetting}
\subsection{Notation and parameter of interest}
\label{notation}
We consider a random sample of \(n\) individuals
\(\{(\Tobs_i,\Etype_i,\Treat_i,\X_i)\}_{i=1}^n\) where \(\Treat\) is a
binary treatment variable assigned at baseline, \(\X\) a
\(d\)-dimensional vector of auxiliary covariates measured at baseline,
\(\Tobs\) a right-censored event time, i.e., \(\Tobs = \T \wedge \C\)
where \(\T\) is the event time, \(\C\) the censoring time, \(\Cause\)
is the event type for which we assume that \(\{\Cause=1\}\) means that
the event of interest occurred and \(\{\Cause=2\}\) that the competing
event occurred, and \(\Etype=\Cause \Ind[\T \leq \C]\) indicates
uncensored observation (we use \(\Ind[\cdot]\) to denote the indicator
function). We assume throughout that \((T,\Cause)\) are conditionally
independent of \(C\) given \((\X,\Treat)\) and that in the case of
tied event and censoring times, i.e., \(\C=\T\), the event time is
earlier. Also, for a fixed time point \(\tau\) we assume that the
probability of right-censoring is bounded away from zero:
\(\Prob[\C>\tau|\Treat,\X]>\epsilon\) where \(\epsilon>0\). We denote
\(\Y(\tau)=\Ind[\T\leq\tau,\Cause=1]\) for the indicator for the event
of interest at time \(\tau\) and note that its expected value is the
absolute risk that the event of interest occurs before time \(\tau\).

To define our target parameter we introduce the potential outcomes
\(\Y^{\treat}(\tau)\), i.e., the response of a randomly selected
individual had that individual, possibly contrary to the fact, been
given treatment \(\Treat=\treat\). Our target parameter is the
expected difference:
\begin{equation*}
\ATE(\tau) = \Esp[\Y^{1}(\tau)-\Y^{0}(\tau)].
\end{equation*}
We make the following assumptions: \(\Y(\tau) = (1-A) \Y^{0}(\tau) +
\Treat \Y^{1}(\tau)\) (consistency assumption), \(\forall a \in \{0,1\}\), \((Y^\treat(\tau),A)\)
are conditionally independent given \(\X\), (no unmeasured
confounders), and \(\forall (\treat,\x) \in \{0,1\} \times \XSupport,
\; \Prob[\Treat=\treat|\x]>0\) (positivity assumption) where
\(\XSupport\subset R^d\) denotes the set of possible values for
\(\X\).

\subsection{Modeling}
\label{sec:orgeacd362}

To estimate the target parameter based on the observed data we
consider the following conditional distributions as nuisance
parameters. The cumulative incidence function \(\Ymodel\) describes
the absolute risk of the event of interest by time \(t\):
\begin{align*}
\Ymodel(t|\Treat,\X)&=\Prob(T \leq t,\Cause=1|\Treat,\X),\\
\intertext{\(\Cmodel\) is the conditional probability of being uncensored}
\Cmodel(t|\Treat,\X)&=\Prob(C> t|\Treat,\X),\\
\intertext{and \(\Tmodel\) describes the propensity of treatment conditional on \X}
\Tmodel(\X)&=\Prob(\Treat=1|\X).
\end{align*}

Under the identifiability assumptions stated in section \ref{notation}
the likelihood of the observed variables
\(\sample_i=(\Tobs_i,\Etype_i,\Treat_i,\X_i)\) factorizes
\citep{begun1983information,gillcar95} and the density of their joint
probability distribution \(P\) with respect to a suitable dominating
measure can be parametrized 
\begin{multline*}
\Prob(\dt,\delta,a,dw)=\{\Cmodel(t-|a,w)\Ymodel(\mathrm d t|a,w)(a\Tmodel(w)+(1-a)(1-\Tmodel(w))H(\mathrm d w)\}^{\Ind[\delta=1]}\\
\{\Cmodel(t-|a,w)\Ftwo(\mathrm d t|a,w)(a\Tmodel(w)+(1-a)(1-\Tmodel(w))H(\mathrm d w)\}^{\Ind[\delta=2]}\\
\{\Smodel(t-|a,w)\Cmodel(\mathrm d t|a,w)(a\Tmodel(w)+(1-a)(1-\Tmodel(w))H(\mathrm d w)\}^{\Ind[\delta=0]}
\end{multline*}
where \(\Ftwo(t|\Treat,\X)=\Prob(T \leq t,\Cause=2|\Treat,\X)\), \(H\)
is the marginal distribution of \(W\), and \(t-\) denotes the
left-handed limit at time \(t\). Our working model for the joint
probability distribution \(P\) leaves the \(H\) part completely
non-parametric but for each of the other nuisance parameters we
specify a (semi-)parametric regression model as our working model and
define a corresponding estimator. Our working model for \(\Ymodel\)
uses the parameterization of \citet{benichou1990estimates} in terms of
the cumulative cause-specific hazard functions \(\Lambda_1\) for the
event of interest and \(\Lambda_2\) for the competing event:
\begin{equation}\label{eq:benichou}
\Ymodel(t|\Treat,\X)= \int_{0}^{t} \Smodel(s-|\Treat,\X) \Lambda_1(\mathrm d s|\Treat,\X) 
\end{equation}
where
\(\Smodel(s|\Treat,\X)=\exp\left\{-(\Lambda_1(s|\Treat,\X)+\Lambda_2(s|\Treat,\X))\right\}\)
is the event free survival function. Specifically we consider two
separate Cox regression models for \(\Lambda_1\) and \(\Lambda_2\)
such that the model is parameterized in terms of the cause-specific
hazard ratios and baseline hazard functions. An alternative
parameterization of \(\Ymodel\) can be obtained by binomial regression
for competing risks \citep{2008_Scheike_Zhang_Gerds} where the Fine-Gray
regression model \citep{fine:gray:1999} is a special case. Our working
models for the censoring mechanism and the propensity of treatment are
a Cox regression model and a logistic regression model,
respectively. Note that all these working models come with their
regular asymptotically linear estimators for the respective nuisance
parameter based on the observed data. Thus, we assume that there exist
regular asymptotically linear estimators
\(\hatYmodel,\hatTmodel,\hatSmodel,\hatCmodel\) with respective large
sample limits \(\Ymodel^*,\Tmodel^*,\Smodel^*,\Cmodel^*\) such that:
\begin{equation}\label{eq:vonmises}
\begin{split}
\sqrt n (\hatTmodel-\Tmodel^*)&= \frac{1}{\sqrt n} \sum_{i=1}^{n} \IFT(\sample_i) + o_p(1),\\
\sqrt n (\hatCmodel-\Cmodel^*)&= \frac{1}{\sqrt n} \sum_{i=1}^{n} \IFC(\sample_i) + o_p(1),\\
\sqrt n (\hatYmodel-\Ymodel^*)&= \frac{1}{\sqrt n} \sum_{i=1}^{n} \IFY(\sample_i) + o_p(1),\\
\sqrt n (\hatSmodel-\Smodel^*)&= \frac{1}{\sqrt n} \sum_{i=1}^{n} \IFS(\sample_i) + o_p(1),
\end{split}
\end{equation}
where \(\sample_i=(\Tobs_i,\Etype_i,\Treat_i,\X_i)\) and
\(\IFY,\IFT,\IFS,\IFC\) are the influence functions corresponding to
the estimators that represent the first order von Mises expansion of
the corresponding statistical functional \citep{vaart98}. If our working
model for \(\Ymodel\) is correctly specified then the asymptotic bias
is zero, \(\Ymodel^*-\Ymodel=0\), and the same holds for the working
models for \(\Tmodel\), \(\Smodel\) and \(\Cmodel\). Note that since
both \(\Ymodel\) and \(\Smodel\) can be expressed as differentiable
functionals of \(\Lambda_j\) for \(j=1,2\), a sufficient condition
for the last two lines of equation \eqref{eq:vonmises} is
\begin{equation*}
\sqrt n (\Lambda_j-\Lambda_j^*)= \frac{1}{\sqrt n} \sum_{i=1}^{n} \IFL[j](\sample_i) + o_p(1),
\end{equation*}
where \(\IFL[j]\) is the influence function of the Cox regression
estimator of the cumulative hazard function \(j\) and \(\Lambda_j^*\)
is the corresponding large sample limit.

In case of a misspecified model, an asymptotic linear expansions of
the estimators as in equation \eqref{eq:vonmises} still continues to
hold under the usual regularity conditions around the least-false
parameters \(\Ymodel^*,\Tmodel^*,\Smodel^*,\Cmodel^*\)
\citep{white,hjort1992,bickel,2001_Gerds_Schumacher}. However, there
would be a large sample bias.

\section{Estimators for the average treatment effect (ATE)}
\label{estimatorATE}
We consider three types of estimators for our estimand \(\ATE(\tau)\).
Each type is based on a different combination of the outcome model, the
treatment model, and the censoring model. We start by defining our
estimators in the case without censoring.

\subsection{Uncensored data}
\label{sec:org13ca819}

The first class of estimators is based on the G-causal parameter
(\cite{robins1986new}, p.1410), also called backdoor adjustment
(\cite{Perl2000}, section 3.2), which yields the G-formula:
\begin{equation*}
\ATE(\tau) = \Esp[\Ymodel(\tau|\Treat=1,\X)) - \Ymodel(\tau|\Treat=0,\X)].
\end{equation*}
Our regression estimator is obtained by substituting \(\hatYmodel\)
for \(\Ymodel\):
\begin{equation}
\widehat{\ATE}_{\text{G-formula}}(\tau) =
 \frac{1}{n} \sum_{i=1}^n \left( \hatYmodel(\tau|\Treat=1,\X_i)
 - \hatYmodel(\tau|\Treat=0,\X_i) \right). \label{eq:ATE:G-formula}
\end{equation}

The second class of estimators uses
inverse probability-of-treatment weights (\(\IPTW)\) and is based on
the formula:
\begin{equation*}
\ATE(\tau) = \Esp\left[\Y(\tau) \left(
\frac{\Treat}{\Tmodel(\X)} 
- \frac{1-\Treat}{1-\Tmodel(\X)}
\right)\right].
\end{equation*}
Our \(\IPTW\) estimator is obtained by substituting \(\hatTmodel\) for
\(\Tmodel\):
\begin{equation}
\widehat{\ATE}_{\IPTW}(\tau) =
 \frac{1}{n} \sum_{i=1}^n \left( 
\Y_i(\tau) \left(
\frac{\Treat_i}{\hatTmodel(\X_i)} 
- \frac{1-\Treat_i}{1-\hatTmodel(\X_i)}
\right)
 \right). \label{eq:ATE:IPTW}
\end{equation}
The third class of estimators combines the
G-formula estimator and the IPTW estimator into a doubly robust
estimator \citep{hernanrobinsbook}. Following
\cite{tsiatis2006semiparametric} (section 13.5) we use the formula
\begin{align*}
\ATE(\tau) &= \Esp\left[ 
\frac{\Y(\tau)\Treat}{\Tmodel(\X)} + \Ymodel(\tau|\Treat=1,\X) \left(1-\frac{\Treat}{\Tmodel(\X)}\right) \right. \notag\\
& \left. \qquad - \frac{\Y(\tau)(1-\Treat)}{1-\Tmodel(\X)} - \Ymodel(\tau|\Treat=0,\X) \left(1-\frac{1-\Treat}{1-\Tmodel(\X)}\right) \right]. \notag 
\end{align*}
Our augmented \(\IPTW\) estimator (denoted \AIPTW) substitutes
\(\hatTmodel\) for \(\Tmodel\) and \(\hatYmodel\) for \(\Ymodel\):
\begin{align}
\widehat{\ATE}_{\AIPTW}(\tau) =& 
\frac{1}{n} \sum_{i=1}^n \frac{\Y_i(\tau)\Treat_i}{\hatTmodel(\X_i)} + \hatYmodel(\tau|\Treat=1,\X_i) \left(1-\frac{\Treat_i}{\hatTmodel(\X_i)} \right) \notag \\
& \qquad - \frac{\Y_i(\tau)(1-\Treat_i)}{1-\hatTmodel(\X_i)} - \hatYmodel(\tau|\Treat=0,\X_i) \left(1-\frac{1-\Treat_i}{1-\hatTmodel(\X_i)} \right). 
\label{eq:ATE:AIPTW}
\end{align}
We refer to \cite{glynn2010introduction} and
\cite{kennedy2016semiparametric} for nice reviews of the doubly robust
\AIPTW estimator in uncensored data.

\subsection{Right-censored data}
\label{sec:org154b644}

In presence of right-censoring, the binary outcome at the time point
of interest \(\Y(\tau)\) is not observed for all subjects, it is only
observed in the event \(\{\C>\T\wedge \tau\}=\{\Tobs >
\tau\}\cup\{\Tobs \leq \tau, \Etype \neq 0\}\). To construct
estimators of the average treatment effect based on the right-censored
data, we combine the estimators of the previous section with inverse
probability-of-censoring weighting (\IPCW) now also using our
estimator \(\hatCmodel\). Note that the G-formula estimator defined in
equation \eqref{eq:ATE:G-formula} does not explicitly involve
\(\Y(\tau)\) and hence can be applied directly in right-censored data
because the outcome model takes care of the censored data. Using that
\(\Ind[\Tobs > \tau]\Y(\tau)=0\), we define the following \IPCW
estimators:
\begin{align}
&\widehat{\ATE}_{\IPTW,\IPCW}(\tau) = \frac{1}{n} \sum_{i=1}^n 
\frac{\Ind[\Tobs_i \leq \tau, \Etype_i \neq 0]}{\hatCmodel(\Tobs_{i}|\Treat_i,\X_i)}
\Y_i(\tau) \left(
\frac{\Treat_i}{\hatTmodel(\X_i)} 
- \frac{1-\Treat_i}{1-\hatTmodel(\X_i)}
\right) \label{eq:ATE:IPTW-IPCW} \\
& \widehat{\ATE}_{\AIPTW,\IPCW}(\tau) = \frac{1}{n} \sum_{i=1}^n 
\frac{\Ind[\Tobs_i \leq \tau, \Etype_i \neq 0]}{\hatCmodel(\Tobs_{i}|\Treat_i,\X_i)} \frac{\Y_i(\tau) \Treat_i}{\hatTmodel(\X_i)} 
+ \hatYmodel(\tau|\Treat=1,\X_i)\left(1-\frac{\Treat_i}{\hatTmodel(\X_i)}\right) \notag \\
& \qquad \qquad - \frac{\Ind[\Tobs_i \leq \tau, \Etype_i \neq 0]}{\hatCmodel(\Tobs_{i}|\Treat_i,\X_i)} \frac{\Y_i(\tau) (1-\Treat_i)}{1-\hatTmodel(\X_i)} - \hatYmodel(\tau|\Treat=0,\X_i) \left(1-\frac{1-\Treat_i}{1-\hatTmodel(\X_i)}\right).  \label{eq:ATE:AIPTW-IPCW}
\end{align}
Both estimators can now be augmented using semi-parametric theory (see
\citet{laan:robins:book03}). In appendix \ref{appendix:Lterm}, we derive the
set of observed-data estimating functions for \(\ATE\). These
estimating equations include an augmentation term which, when set to
0, leads to the \(\IPCW\) estimators (equations \eqref{eq:ATE:IPTW-IPCW}
and \eqref{eq:ATE:AIPTW-IPCW}). Alternatively the augmentation term can
be chosen in order to minimize the asymptotic variance of the
corresponding estimator. This choice lead to the following estimators
(see appendix \ref{appendix:Lterm} for details):
\begin{align}
&\widehat{\ATE}_{\IPTW,\AIPCW}(\tau) = \widehat{\ATE}_{\IPTW,\IPCW}(\tau)
 + \frac{1}{n} \sum_{i=1}^n \hat{I}(\Tobs_i,\tau|\Treat_i,\X_i) \left(\frac{\Treat_i}{\hatTmodel(\X_i)} - \frac{1-\Treat_i}{1-\hatTmodel(\X_i)}\right)  \label{eq:ATE:IPTW-AIPCW} \\
&\widehat{\ATE}_{\AIPTW,\AIPCW}(\tau)= \widehat{\ATE}_{\AIPTW,\IPCW}(\tau) 
 + \frac{1}{n} \sum_{i=1}^n \hat{I}(\Tobs_i,\tau|\Treat_i,\X_i) \left(\frac{\Treat_i}{\hatTmodel(\X_i)} - \frac{1-\Treat_i}{1-\hatTmodel(\X_i)}\right)  \label{eq:ATE:AIPTW-AIPCW} \\
&\text{where } \hat{I}(\Tobs_i,\tau|\Treat_i,\X_i) = \int_0^{\Tobs_i \wedge \tau} \frac{\hatYmodel(\tau|\Treat_i,\X_i)-\hatYmodel(t|\Treat_i,\X_i)}{\hatSmodel(t|\Treat_i,\X_i)} \frac{1}{\hatCmodel(t|\Treat_i,\X_i)} d\hat{M}_{i}^{\C}(t). \notag
\end{align}
Here \(N_i^C(t) =\Ind[\Tobs_i \leq t, \Etype_i = 0]\) denotes the
censoring counting process of subject \(i\) and \(\Lambda^C\) the
 cumulative hazard function of \(\Cmodel\) such that
 \(M_i^C(t)= N_i^C(t) - \int_0^t \Ind[\tilde T_i \geq s]
 \Lambda^C(\mathrm d s|\Treat_i,\X_i)\) is a 0 mean
 process
\citep[a martingale with respect ot the natural filtration, see for example][section II.4]{abgk}.
We use the notation \(\hat{M}_{i}^{\C}(t)=N_i^C(t) -
\int_0^t \Ind[\tilde T_i \geq s]
 \hat\Lambda^C(\mathrm d s|\Treat_i,\X_i)\) and \(M_{i}^{\C,*}(t)=
N_i^C(t) - \int_0^t \Ind[\tilde T_i \geq s]
 \Lambda^{C,*}(\mathrm d s|\Treat_i,\X_i)\) where
\(\Lambda^{C,*}\) is the large sample limit of \(\hat{\Lambda}^{\C}\).

\section{Asymptotic properties}
\label{asymptotic}
In this section, we study the asymptotic properties of the following
estimators: \(\widehat{\ATE}_{\text{G-formula}}(\tau)\),
\(\widehat{\ATE}_{\IPTW,\IPCW}(\tau)\), and
\(\widehat{\ATE}_{\AIPTW,\AIPCW}(\tau)\).

\subsection{Consistency}
\label{sec:orga76c23a}

By equation \eqref{eq:vonmises} and the law of large numbers we have
\begin{equation*}
\lim_{n\to\infty}\widehat{\ATE}_{\text{G-formula}}(\tau)=\Esp[\Ymodel^*(\tau|\Treat=1,\X)) - \Ymodel^*(\tau|\Treat=0,\X)].
\end{equation*}
Thus, if the outcome model is correctly specified at \(\tau\), i.e.,
for \(\treat\in\{0,1\}\) and almost all \(w\) 
\(\Ymodel(\tau|a,w)=\Ymodel^*(\tau|a,w)\), then
\(\widehat{\ATE}_{\text{G-formula}}\) is a consistent estimator for
\(\ATE(\tau)\). Similarly, we have under the assumptions of Section
\ref{CRsetting}
\begin{multline*}
\lim_{n\to\infty}\widehat{\ATE}_{\IPTW,\IPCW}(\tau)=
\Esp\Big[\frac{G(\Tobs|\Treat,\X)}{G^*(\Tobs|\Treat,\X)}
\Big\{
  \frac{\Ymodel(\tau|\Treat=1,\X)\Tmodel(\X)}{\Tmodel^*(\X)}.\\
- \frac{\Ymodel(\tau|\Treat=0,\X)(1-\Tmodel(\X))}{1-\Tmodel^*(\X)}
\Big\}
\Big]
\end{multline*}
Hence, if the working models for the treatment and the censoring
mechanism are correctly specified, i.e., \(\Tmodel(w)=\Tmodel^*(w)\) and
\(\Cmodel(s|a,w)=\Cmodel^*(s|a,w)\) for all \(s\in[0,\tau]\), \(a\in\{0,1\}\) and almost
all \(w\), then \(\widehat{\ATE}_{\IPTW,\IPCW}(\tau)\) is
consistent. The following theorem states sufficient conditions under
which \(\widehat{\ATE}_{\AIPTW,\AIPCW}\) is consistent.
\begin{theorem}
Under the assumptions stated in Section \ref{CRsetting}, the estimator
\(\widehat{\ATE}_{\AIPTW,\AIPCW}(\tau)\) is consistent whenever one of the following conditions is satisfied
 for all \(s\in[0,\tau]\), \(a\in\{0,1\}\) and almost all \(w\):
\begin{enumerate}
\item \(\Cmodel^*(s|a,w)=\Cmodel(s|a,w)\) and \(\Ymodel^*(\tau|a,w)=\Ymodel(\tau,a,w)\) 
\item \(\Cmodel^*(s|a,w)=\Cmodel(s|a,w)\) and \(\Tmodel^*(w)=\Tmodel(w)\)
\item \(\Ymodel^*(s|a,w)=\Ymodel(s|a,w)\) and \(\Smodel^*(s|a,w)=\Smodel(s|a,w)\) 
\end{enumerate}
\end{theorem}

Proof: Roughly, when the censoring model is correctly
specified, 1. and 2. follow from the fact that
\(\widehat{\ATE}_{\AIPTW,\AIPCW}(\tau)\) and
\(\widehat{\ATE}_{\AIPTW}(\tau)\) have the same large sample
limit. When the censoring model is misspecified but the outcome and
survival models are correctly specified then
\(\widehat{\ATE}_{\AIPTW,\AIPCW}(\tau)\) and
\(\widehat{\ATE}_{\text{G-formula}}(\tau)\) have the same large sample
limit, which gives 3. Appendix \ref{appendix:consistency} provides the
details.
\subsection{Asymptotic distribution}
\label{distributionATE}
All estimators described in the previous section can be written as 
averages of the estimated nuisance parameters:
\begin{align*}
\widehat{\ATE}_x(\tau) =& \frac{1}{n} \sum_{i=1}^{n} h_{x}(\tau;\sample_i;\hatYmodel,\hatTmodel,\hatSmodel,\hatCmodel) \\
\text{where } x \in& \{{\scriptsize\text{G-formula}};{\scriptsize\text{IPTW,IPCW}};{\scriptsize\text{AIPTW,IPCW}}; {\scriptsize\text{IPTW,AIPCW}}; {\scriptsize\text{AIPTW,AIPCW}}\} 
\end{align*}
and a suitable function \(h_{x}\). For instance,
\begin{equation*}
h_{\text{G-formula}}(\tau;\sample_i;\hatYmodel,\hatTmodel,\hatSmodel,\hatCmodel)=
\hatYmodel(\tau|\Treat=1,\X_i) -
\hatYmodel(\tau|\Treat=0,\X_i). 
\end{equation*}
If the nuisance parameters were known, say equal to
\((\Ymodel^*,\Tmodel^*,\Smodel^*,\Cmodel^*)\), the correspondingly
defined plug-in estimators would be simple averages of independent and
identically distributed quantities with influence function:
\begin{equation}
\widetilde\IF_{x}(\tau; \sample_i) = h_{x}(\tau;\sample_i;\Ymodel^*,\Tmodel^*,\Smodel^*,\Cmodel^*) - \ATE_x^*(\tau) \label{eq:IFnoNuisance}
\end{equation}
where \(\ATE_x^*\) is the large sample limit of \(\widehat{\ATE}_x\).
From the central limit theorem, we would get that the estimators are
asymptotically normal with asymptotic variance equal to the 
variance of the influence function. However, in practice the nuisance
parameters are estimated with the same data and the asymptotic
expansions of the estimators of the average treatment effect involve
the influence functions of the estimators of the nuisance parameters
given in equation \eqref{eq:vonmises}. The general idea is to apply the
functional delta method (\cite{vaart98}, chapter 20) to obtain a von
Mises expansion of the form:
\begin{equation*}
\sqrt n (\widehat{\ATE}_x(\tau)-\ATE^*_x(\tau))=\frac{1}{\sqrt n} \sum_{i=1}^{n} \IF_x(\tau;\sample_i) + o_P(1).
\end{equation*}
The influence function has two terms:
\begin{equation}
\IF_{x}(\tau; \sample_i) =
\widetilde\IF_{x}(\tau; \sample_i)
+ \phi_x(\tau;\sample_i;\IFY,\IFT,\IFS,\IFC)
\label{eq:IFnuisance}
\end{equation}
where a function \(\phi_x\) (the derivate of \(h_x\)) relates to the
influence functions of the estimators of the nuisance parameters. In
the case of the G-formula estimator,
\begin{multline*}
\phi_{\text{G-formula}}(\tau;\sample_i;\IFY,\IFT,\IFS,\IFC) = \Esp[\IFY(\tau,A=1,\X;\sample_i)|\sample_i]  \\
-\Esp[\IFY(\tau,A=0,\X;\sample_i)\big |\sample_i]
\end{multline*}
and for the \(\IPTW,\IPCW\) estimator:
\begin{align}
&\phi_{\IPTW,\IPCW}(\tau;\sample_i;\IFY,\IFT,\IFS,\IFC) \notag \\
=& -\Esp\left[ \IFT(\X;\sample_i) \frac{\Ind[\Tobs \leq \tau,\Etype \neq 0]}{\Cmodel^*(\Tobs| \Treat,\X)} \Y(\tau) \left(\frac{\Treat}{\Tmodel^*(\X)^2}+\frac{1-\Treat}{\left(1-\Tmodel^*(\X)\right)^2}\right)\Bigg |\sample_i \right] \label{eq:iid-IPTW-IPCW-T} \\
&  -\Esp\left[ \IFC(\Tobs,\Treat, \X;\sample_i) \frac{\Ind[\Tobs \leq \tau,\Etype \neq 0]}{\Cmodel^*(\Tobs | \Treat,\X)^2} \Y(\tau) \left(\frac{\Treat}{\Tmodel^*(\X)}-\frac{1-\Treat}{1-\Tmodel^*(\X)}\right) \Bigg |\sample_i \right]. \label{eq:iid-IPTW-IPCW-C}
\end{align}
The formula for the influence function of \(\widehat{\ATE}_{\AIPTW,\AIPCW}\) is more complex and can be found
in Appendix \ref{appendix:iid}. 

Under the assumptions stated in Section \ref{CRsetting}, and in
particular under equation \eqref{eq:vonmises}, the functional delta
method yields that the asymptotic distribution of the estimator
\(\widehat{\ATE}_{x}\) is a normal
distribution with variance equal to the variance of the
influence function. The variance of \(\widehat{\ATE}_{x}\) can then be
estimated based on an estimate \(\widehat\IF_{x}\) of the influence
function:
\(\frac 1 n\sum_{i=1}^n \big(\widehat\IF_{x}(\sample_i)\big)^2\).

\textbf{Remark 1}: In appendix \ref{appendix:iid} we show that when all working
models are correctly specified, then we have
\(\phi_{\AIPTW,\AIPCW}(\tau;\sample_i;\IFY,\IFT,\IFS,\IFC) = 0\). In
this case a consistent estimator of the asymptotic variance of
\(\widehat{\ATE}_{\AIPTW,\AIPCW}(\tau)\) is given by 
\begin{multline*}
\frac{1}{n} \sum_{i=1}^n \left( \hatYmodel(\tau|\Treat = 1, \X_i) - \hatYmodel(\tau|\Treat = 0, \X_i) - \hat{\ATE}_{\AIPTW,\AIPCW}(\tau) \right.  \\
\left. + \left( \frac{\Treat_i}{\hatTmodel(\X_i)} - \frac{1-\Treat_i}{1-\hatTmodel(\X_i)} \right) \left(\frac{\Ind[\Tobs_i \leq \tau,\Etype_i \neq 0]Y_i(\tau)}{\hatCmodel(\Tobs_i| \Treat_i,\X_i)}  - \hatYmodel(\tau|\Treat_i,\X_i) + \hat{I}(\Tobs_i,\tau|\Treat_i,\X_i) \right) \right)^2.
\end{multline*}
This result is a consequence of the orthogonality between the
estimating function and the nuisance parameter tangent space, see also
\citet[][Remark 4, Section 3.3]{tsiatis2006semiparametric}.

\section{Empirical studies}
\label{simulation}
The following simulation studies investigate the bias-variance
tradeoff of the various estimators under model misspecification and
the small sample coverage based on the asymptotic variance formula.

\subsection{Simulation setting}
\label{sec:orgc57cfb6}
\begin{figure}[htbp]
\centering
\includegraphics[width=0.7\textwidth]{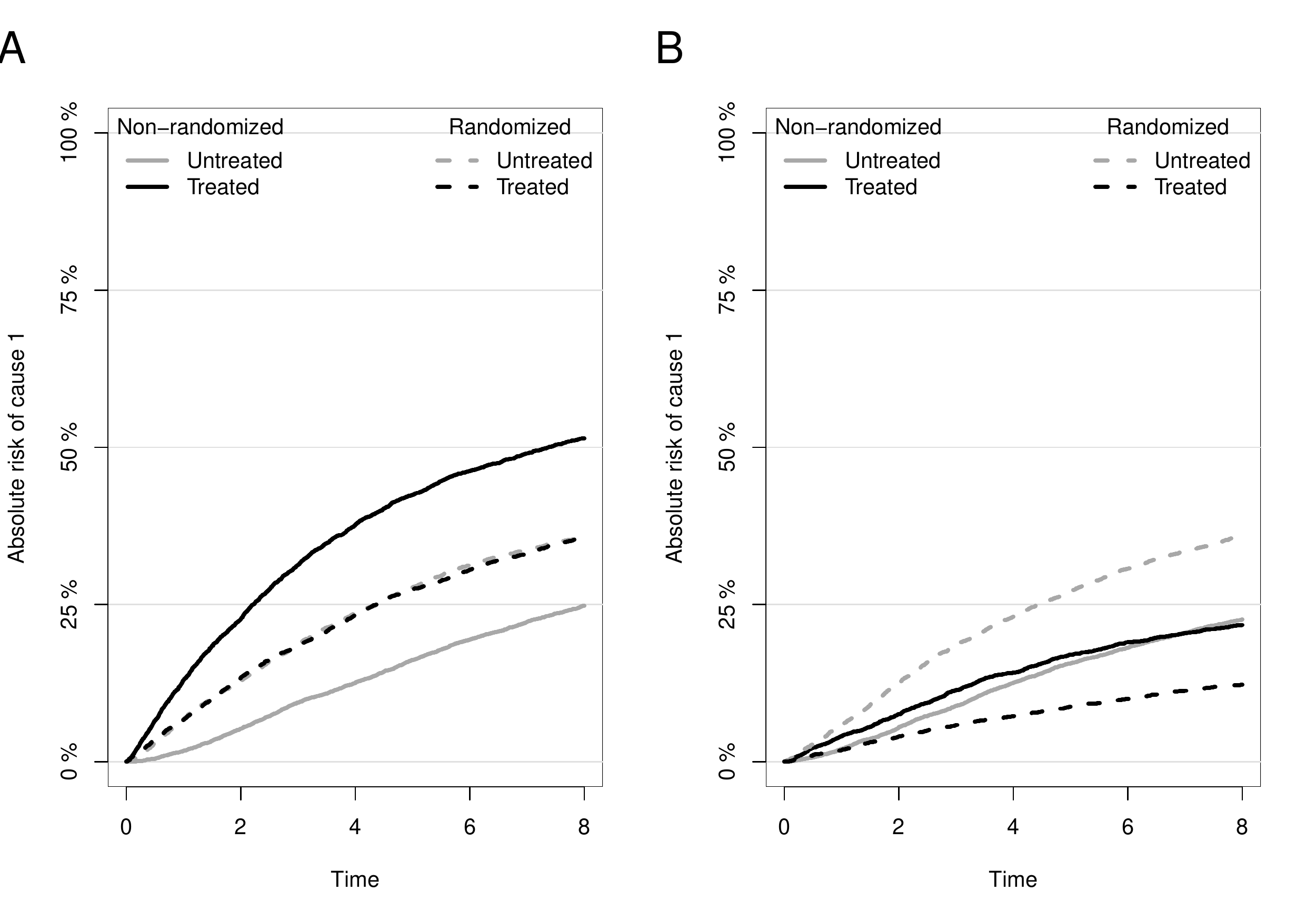}
\caption{\label{fig:sim-set}
Illustration of the data generating mechanism used in our simulation studies. Shown are the Aalen-Johansen estimates for the absolute risks of cause 1 in both treatment arms in two independently drawn datasets (non-randomized and randomized) each of size n=10,000. Panel A: The treatment effect is zero. In the non-randomized world, the Aalen-Johansen estimate of the 8-year risk difference is large. Panel B: The treatment has a protective effect. In the non-randomized world, the Aalen-Johansen estimate of the 8-year risk difference is about zero.}
\end{figure}

In total, 12 auxiliary covariates are simulated, 6 having a standard
normal distribution (\(\X_1,\ldots,\X_6\)) and the remaining 6 having a
Bernoulli distribution (\(\X_7,\ldots,\X_{12}\)). A binary treatment
variable is drawn following a logistic regression model. We use three
Cox-Weibull regression models \cite[Table II,][]{Bender2005} to
simulate three latent times conditional on treatment and auxiliary
covariates, one for the event of interest, one for the competing risk
and one for the right-censoring time. The observed time is then
obtained as the minimum of the three latent times and the event status
corresponds to the event with the smallest latent time. In the main
analyses the 12 auxiliary covariates are independent. The covariate
effects on the treatment, hazard rate of the event of interest, the
hazard rate of the competing risk and the hazard rate of the censoring
are controlled by including additive effects of the 6 binary
variables, the 6 continuous variables and the squares of the 6
continuous variables into the linear predictors of the logistic
regression and the Cox-Weibull regression models, respectively. The
effect of treatment on the three hazard rates is controlled by three
additional regression parameters. Note that the randomized world
corresponds to setting all regression parameters of the logistic
regression model to zero and deviations from the randomized world
can be controlled by varying these covariate effects (Figure
\ref{fig:sim-set}).

For various parameter settings we report results of the estimators
G-formula (equation \ref{eq:ATE:G-formula}), \(\IPTW,\IPCW\) (equation
\ref{eq:ATE:IPTW-IPCW}), and \(\AIPTW,\AIPCW\) (equation
\ref{eq:ATE:AIPTW-AIPCW}) across 1,000 simulated datasets. These
estimators are implemented in R \citep{Rlang} in the package
\texttt{riskRegression} (\cite{riskRegression}, function \texttt{ateRobust}). When
estimating the variance of the \(\AIPTW,\AIPCW\) estimators, we
consider two estimators for the influence function. The first, denoted
\(\tilde{\IF}_{\AIPTW,\AIPCW}\), only estimates the first term of
equation \eqref{eq:IFnuisance} since the second term is 0 in correctly
specified models. The second estimates both terms and is denoted
\(\IF_{\AIPTW,\AIPCW}\). However we have not implemented all the terms
necessary to compute \(\phi_{\AIPTW,\AIPCW}\): the current
implementation is equivalent to neglecting the uncertainty relative to
the censoring weights and the augmentation term
\(I(\Tobs,\tau|\Treat,\X)\). The R-code of our simulation studies is
available as supplementary material.

\subsection{Simulation results}
\label{sec:org624b155}

\begin{figure}[htbp]
\centering
\includegraphics[width=\textwidth]{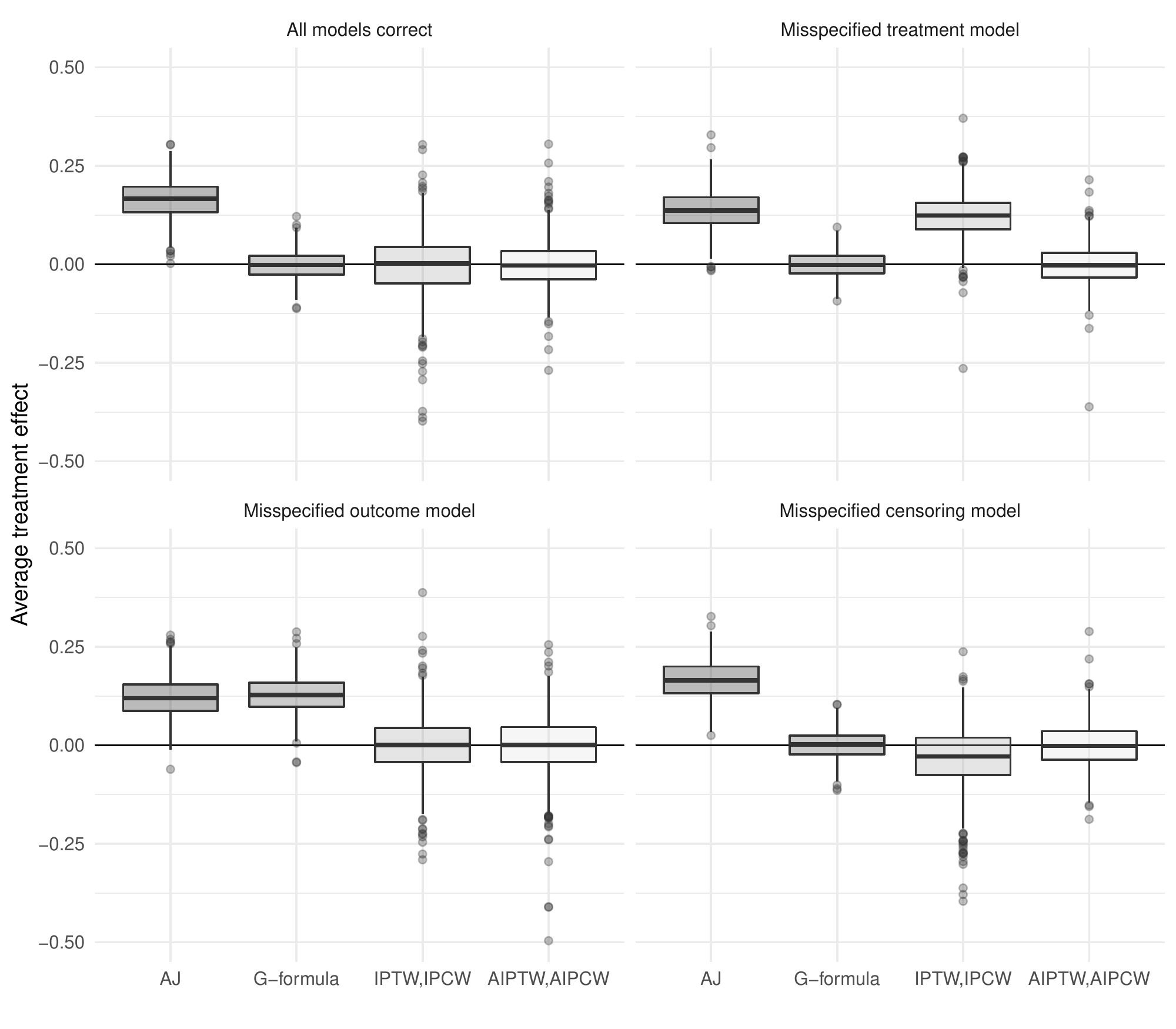}
\caption{\label{fig:sim-bias}
Simulation setting where there is no treatment effect (panel A of Figure \ref{fig:sim-set}). Boxplots show results of 1000 simulated data sets (each with sample size 500) and each of 4 methods for estimating the average 10-year risk difference between treated and untreated subjects. Upper left panel: all regression models (treatment, event of interest, competing risk, censoring) are correctly specified. Upper right panel: the treatment model is misspecified (missing covariates and missing quadratic effects). Lower left panel: the event of interest and the competing risk models are misspecified (missing covariates and missing quadratic effects). Lower right panel: the censoring model is misspecified  (missing quadratic effects).}
\end{figure}

\begin{figure}[htbp]
\centering
\includegraphics[width=\textwidth]{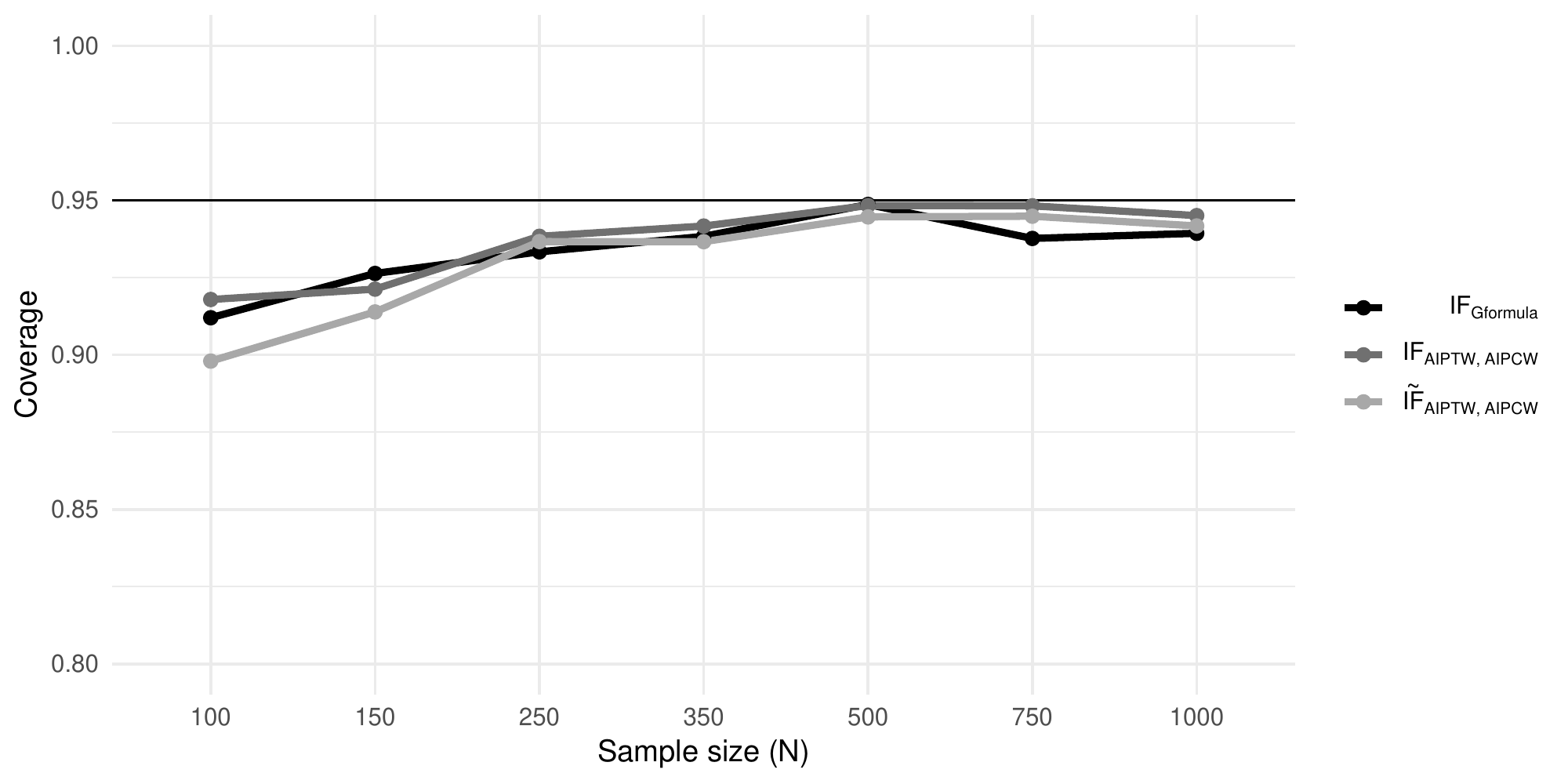}
\caption{\label{fig:sim-N}
Effect of sample size on coverage in a simulation setting where there is no treatment effect (panel A of Figure \ref{fig:sim-set}). The black curve corresponds to the G-formula estimator, the dark grey curve to the \(\AIPTW,\AIPCW\) estimator using the full influence function to estimate the variance, and the light gray curve to the \(\AIPTW,\AIPCW\) estimator using only the first term of the influence function to estimate the variance.}
\end{figure}

We report results for a data generating model without treatment effect
(Panel A, Figure \ref{fig:sim-set}). The figure shows Aalen-Johansen
estimates \citep{aalenjohansen78,abgk} of the cumulative incidence
functions. Similar results are obtained when considering a non-zero
treatment effect but then the ``true'' value needs to be obtained
empirically. Model misspecification is simulated by omitting
covariates and quadratic effects. We created four scenarios. In the
first one, all models are correctly specified. In the three other
scenarios, precisely one of the censoring, outcome, or treatment
models is misspecified.  As shown in figure \ref{fig:sim-bias} (upper
panel), the AIPTW,AIPCW estimator is consistent even when one of the
models (outcome, treatment, or censoring) is misspecified. The
G-formula estimator and the IPTW,IPCW estimator need one or two models
to be correctly specified to be consistent - the outcome model for the
G-formula estimator and both the treatment and censoring models for
the IPTW,IPCW estimator. The G-formula estimator appears to be less
variable compared to the other estimators. The IPTW,IPCW estimator is
at least as variable but often more variable than the AIPTW,AIPCW
estimator. The coverage of the G-formula estimator and AIPTW,AIPCW
estimator is found satisfactory even in small samples when the outcome
model is correctly specified (Figure \ref{fig:sim-N}).

\section{Real data application}
\label{application}
\begin{figure}[htbp]
\centering
\includegraphics[width=0.7\textwidth]{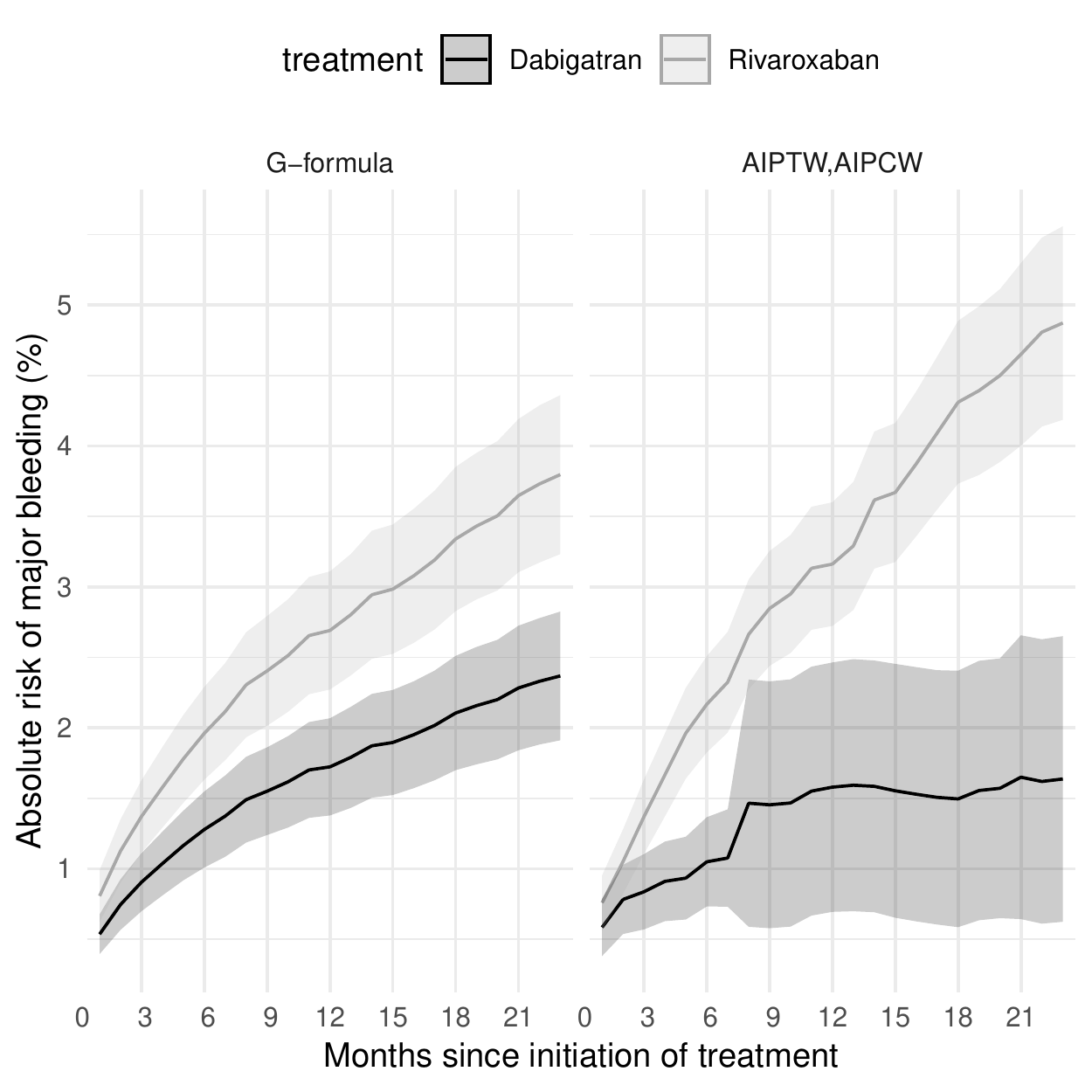}
\caption{\label{fig:illu}
Illustration in data of Danish registry study. Lines are absolute risk estimates with time-pointwise 95\% confidence limits using G-formula (left panel) and AIPTW,AIPCW (right panel).}
\end{figure}

For the sole purpose of illustration, we consider a subset of the data
presented in \citet{staerk2018}. This Danish registry study included
n=21149 patients with a diagnosis of atrial fibrillation (AF) in the
period 2012-2016 who initiated anticoagulation treatment with a
standard dose of dabigatran (n=7078) or rivaroxaban (n=6868) or
apixaban (n=7203). All three treatments belong to the group of
non-vitamin K antagonist oral anticoagulants (NOAC's). Here we
consider only data from patients that initiated treatment with either
dabigatran or rivaroxaban. The follow up started at the date of
treatment initiation. The original study \citet{staerk2018} presented
results on several adverse endpoints including thromboembolism/stroke
and major bleeding where death without the endpoint is the only
competing risk. Here we consider the analysis of the endpoint major
bleeding where death without major bleeding or shift or
discontinuation of treatment are the competing risk. The treatment
assignment is not randomized but there are official guidelines and
presumably also doctor preferences which most likely also depend on
the patient characteristics. Note that the results presented here for
G-formula are not directly comparable to those presented in
\citet{staerk2018} because we here restrict all Cox regression models to
the subset of the dabigatran and rivaroxaban patients. Otherwise we
use the same covariate adjustment as described in detail in
\citet{staerk2018} for all Cox regression models and for the logistic
regression model of the treatment mechanism. Figure \ref{fig:illu}
displays the estimates absolute risk of major bleeding obtained with
G-formula and AIPTW,AIPCW. Within the limitation of the available confounder
information the results can be interpreted as what one would have
observed in a hypothetical world where all patients initiated
dabigatran (or rivaroxaban), respectively.

The interpretation of these results is limited to the population of
patients who initiated either dabigatran or rivaroxaban in the period
2012-2016. Based on the AIPTW,AIPCW estimate evaluated at 12 months,
the interpretation could be as follows. If every patient had received
dabigatran the 1-year risk [95\% confidence interval] of a major
bleeding would have been 1.58\% [0.60;2.57] lower compared to when
every patient had received rivaroxaban.  Interestingly, the
AIPTW,AIPCW estimates of the risk differences are larger in magnitude
compared to the G-formula estimates. For example, the estimate of ATE
(12-month) using G-formula is only 0.97\% [0.40;1.54].

\section{Discussion}
\label{sec:org86b9f75}

In presence of completely observed outcomes, estimation of the average
treatment effect can be performed using estimators based on the
G-formula, inverse probability of treatment weighting (IPTW), or a
combination of both (AIPTW). While these are classical tools in causal
inference (see for e.g., part 2 of \cite{hernanrobinsbook}), we review
in this article their extension to right-censored data and to the
presence of competing risks. Using results from semi-parametric
theory, we derive the augmentation term relative to the working model
for the censoring distribution. We investigate the robustness of this
new estimator against misspecification of the working models. We also
show the asymptotic normality of this estimator and derive an
analytical formula for its influence function which can be used to
estimate the variance of the estimator. The variance of our estimator
may depend on the estimators of the nuisance parameters. In our
software implementation \citep{riskRegression}, we focus on the use of
cause-specific Cox regression models for the outcome model, a logistic
regression for the treatment model, and a Cox regression model for the
censoring model. An alternative would be to use a Fine-Gray regression
model for the outcome. However, then one would need an additional
working regression model for the conditional event-free survival
function \(\Smodel(\cdot|\Treat,\X)\). To simplify the implementation,
we currently do not estimate the variability related to the estimation
of the censoring distribution \(\Cmodel(\cdot|\Treat,\X)\) and the
augmentation term \(I(\cdot,\tau|\Treat,\X)\). In to our simulation
study this omission did not have a large effect and the coverage of
our confidence intervals was sufficient.

The simulations confirm the superiority of the \(\AIPTW,\AIPCW\)
estimator over the \(\IPTW,\IPCW\) estimator. They also show that the
G-formula estimator is less variable than the \(\AIPTW,\AIPCW\)
estimator when the outcome model is correctly specified. However, the
G-formula estimator has a bias that the \(\AIPTW,\AIPCW\) estimator
does not have when the outcome model is misspecified. It is worth
noting that the definition of the G-formula estimator is unchanged in
presence of censoring - only the outcome model has to properly handle
censoring.

Competing risks essentially lead to a change of the definition of the
outcome, where we use \(\Ind[T \leq \tau, \Cause = 1]\) instead of
\(\Ind[T \leq \tau]\). However one should not overlook that the
presence of competing risks complexifies the assessment of the
treatment effect, especially when the treatment has a positive effect
on the cause of interest but a negative effect on the competing
events. We refer to \cite{young2018choice} for a detailed discussion of
the implications of how the estimand is defined in presence of
competing risks.

Recently, \cite{lesko2017bias} pointed out that bias will occur if we do
not have the correct models for the probability of the outcome of
interest \(\Ymodel(\tau|\Treat,\X)=\Prob[\T\leq \tau,
\Cause=1|\Treat,\X]\), in particular when the model for the hazard rate
of the competing risk \(\Lambda_2\) is misspecified. In practice this
means that, if we estimate the outcome model via a cause-specific Cox
regression models, both conditional hazard functions need to be
correctly specified. While our approach relies on prior knowledge to
define the working models, automated techniques and the use of
cross-validation \citep{benkeser2018estimating} may be preferable when
prior knowledge is sparse. Indeed, the Cox regression model makes the
assumption of proportional hazards which may not always be
appropriate. This assumption can be relaxed, e.g., by using stratified
baseline hazard functions, time varying coefficients
\citep{martinussen2007dynamic}, or an alternative approach that does not
rely on this assumption (e.g., using pseudo-observations
\citep{andersen2017causal}).

We have focused on a binary treatment variable. In the case of a
multi-valued treatment variable the several estimands can be defined
depending on the type of the treatment variable (ordinal versus
nominal), see \cite{imbens2000role} for a nice discussion. One option is
to compare each pair of treatments in the subpopulation of subjects
treated with either of the treatments. This is what we have done in
our real data analysis.

We have also focused on a single time point to evaluate the treatment
effect. However, our methods can be extended to multiple time points,
perhaps at the cost of a multiple testing issue.

Handling time-varying treatments and therefore
possible time-varying confounding is more challenging and beyond the
scope of this article; we refer the refer the interested reader to
\citep{bekaert2010adjusting,daniel2013methods,moodie2014marginal,hernanrobinsbook}.

\section{Acknowledgment}
\label{sec:org7a2a9e1}
\begin{acknowledgement}
B.M.H.O. was supported by  Hjerteforeningen Forskningsst\o{}tte (nr. 15-R99-A5954 015-S15), the Lundbeck foundation (R231-2016-3236) and Marie-Curie-NEUROMODEL (746850).
\end{acknowledgement}
\section{Conflict of interest}
\label{sec:org37c683d}
The authors have declared no conflict of interest.

\bibliographystyle{chicago}
\bibliography{main-competing-risk-ate}

\renewcommand{\thesection}{}
\section{Appendix}
\label{sec:org5b113ec}
\renewcommand{\thesubsection}{\Alph{subsection}}

\subsection{Estimating equation for the \AIPTW,\AIPCW estimator}
\label{appendix:Lterm}
For a generic individual with full data \(\sample\), we denote by
\(\Score_{\AIPTW}\) the estimating function:
\begin{align*}
\Score_{\AIPTW}(\tau;\sample) =& \, \Y(\tau) \left(
\frac{\Treat}{\Tmodel(\X)} 
- \frac{1-\Treat}{1-\Tmodel(\X)} 
\right) + \Ymodel(\tau|\Treat=1,\X) \left(1-\frac{\Treat}{\Tmodel(\X)}
 \right) \\
&  -  \Ymodel(\tau|\Treat=0,\X) \left(1-\frac{1-\Treat}{1-\Tmodel(\X)}
 \right) - \ATE(\tau).
\end{align*}
Semi-parametric theory (e.g., \citep{tsiatis2006semiparametric} - chapter
9, formula 9.34) gives the following augmented estimating equation:
\begin{equation*}
\sum_{i=1}^n \frac{\Ind[\C_i > \T_i \wedge \tau]\Score_{\AIPTW}(\tau;\sample_i)}{\Cmodel(\Tobs_i \wedge \tau|\Treat_i,\X_i)} 
 - \int_0^{\tau \wedge \Tobs_i} \frac{f_{\AIPTW}(t,\sample_i)}{\Cmodel(t|\Treat_i,\X_i)} dM_{i}^{\C}(t) = 0
\end{equation*}
where \(f_{\AIPTW}\) is an element of the space of real valued
functions well-defined on the support of
\((\Tobs,\Etype,\Treat,\X)\). To fully define the estimating equation
it remains to define what is \(f_{\AIPTW}\). It is reasonable to
choose \(f_{\AIPTW}\) such the estimator has the smallest asymptotic
variance, i.e., its influence function has the smallest
variance. Theorem 10.1 and 10.4 in \citep{tsiatis2006semiparametric}
gives that this is achieved by taking \(f_{\AIPTW}(t;\sample) =
-\Esp[\Score_{\AIPTW}(\tau;\sample)|\T > t,\Treat,\X]\). It follows that:
\begin{align*}
f_{\AIPTW}(t;\sample) =& -\left(\Esp\left[ \Y(\tau) | T>t,\Treat,\X\right] \left(
\frac{\Treat}{\Tmodel(\X)} 
- \frac{1-\Treat}{1-\Tmodel(\X)} 
\right) - \ATE(\tau) \right. \\
&  \left. + \Ymodel(\tau|\Treat=1,\X) \left(1-\frac{\Treat}{\Tmodel(\X)} \right) -  \Ymodel(\tau|\Treat=0,\X) \left(1-\frac{1-\Treat}{1-\Tmodel(\X)}
 \right) \right) ,
\end{align*}
where
\begin{align*}
\Esp[Y(\tau)|T>t,\Treat,\X] &= \Prob[\T\leq\tau,\Etype=1|\T>t,\Treat,\X] = \frac{\Prob[t<\T\leq\tau,\Etype=1|\Treat,\X]}{\Prob[\T>t|\Etype,\X]} \\
&= \frac{\Prob[\T\leq\tau,\Etype=1|\Treat,\X]-\Prob[\T\leq t,\Etype=1|\Treat,\X]}{\Prob[\T>t|\Treat,\X]} \\
&= \frac{\Ymodel(\tau|\Treat,\X)-\Ymodel(t|\Treat,\X)}{\Smodel(t|\Treat,\X)}.
\end{align*}
Since:
\begin{align}
\int_0^{\tau \wedge \Tobs} \frac{1}{\Cmodel(t|\Treat,\X)} dM^{\C}(t) &= \int_0^{\tau \wedge \Tobs} \exp(\Lambda^{\C}(t|\Treat,\X)) d(N_{\C}(t)-\Lambda^{\C}(t|\Treat,\X)) \notag \\
&= 1-\frac{\Ind[\C > \Tobs \wedge \tau]}{\Cmodel(\Tobs \wedge \tau|\Treat,\X)}, \label{eq:magic}
\end{align}
we obtain the augmented estimating equation for the \(\AIPTW\) estimator:
\begin{align*}
0 = & \sum_{i=1}^n \frac{\Ind[\C_i > \T_i \wedge \tau]}{\Cmodel(\Tobs_i \wedge \tau|\Treat_i,\X_i)}
\left(\Y_i(\tau)\left( \frac{\Treat_i}{\Tmodel(\X_i)} - \frac{1-\Treat_i}{1-\Tmodel(\X_i)}\right)
 + \Ymodel(\tau|\Treat=1,\X_i) \left(1-\frac{\Treat_i}{\Tmodel(\X_i)} \right) \right. \\
& \qquad \qquad \qquad \qquad \qquad \left. -  \Ymodel(\tau|\Treat=0,\X_i) \left(1-\frac{1-\Treat_i}{1-\Tmodel(\X_i)} \right)
- \ATE(\tau) \right)  \\
& \qquad + \left(\frac{\Treat_i}{\Tmodel(\X_i)} - \frac{1-\Treat_i}{1-\Tmodel(\X_i)}\right) \int_0^{\tau \wedge \Tobs_i} \frac{\Ymodel(\tau|\Treat_i,\X_i)-\Ymodel(t|\Treat_i,\X_i)}{\Smodel(t|\Treat_i,\X_i)} \frac{1}{\Cmodel(t|\Treat_i,\X_i)} dM_{i}^{\C}(t)  \\
& \qquad + \left(\Ymodel(\tau|\Treat=1,\X_i) \left(1-\frac{\Treat_i}{\Tmodel(\X_i)}\right) - \Ymodel(\tau|\Treat=0,\X_i) \left(1-\frac{1-\Treat_i}{1-\Tmodel(\X_i)} \right) - \ATE(\tau)\right) \\
& \qquad\qquad\qquad \left(1-\frac{\Ind[\C_i > \T_i \wedge \tau]}{\Cmodel(\Tobs_i \wedge \tau|\Treat_i,\X_i)} \right),
\end{align*}
i.e. denoting \(I(\Tobs_i,\tau|\Treat_i,\X_i)=\int_0^{\tau \wedge
\Tobs_i}
\frac{\Ymodel(\tau|\Treat_i,\X_i)-\Ymodel(t|\Treat_i,\X_i)}{\Smodel(t|\Treat_i,\X_i)}
\frac{1}{\Cmodel(t|\Treat_i,\X_i)} dM_{i}^{\C}(t)\):
\begin{align}
0 =& \sum_{i=1}^n  \left(
\frac{\Treat_i}{\Tmodel(\X_i)} 
- \frac{1-\Treat_i}{1-\Tmodel(\X_i)}
 \right) \left(\frac{\Y_i(\tau)\Ind[\C_i > \T_i \wedge \tau]}{\Cmodel(\Tobs_i \wedge \tau|\Treat_i,\X_i)} + I(\Tobs_i,\tau|\Treat_i,\X_i) \right) \notag \\
&+ \Ymodel(\tau|\Treat=1,\X_i) \left(1-\frac{\Treat_i}{\Tmodel(\X_i)}
 \right) -  \Ymodel(\tau|\Treat=0,\X_i) \left(1-\frac{1-\Treat_i}{1-\Tmodel(\X_i)}
 \right) - \ATE(\tau). \label{eq:EE:AIPTW-AIPCW}
\end{align}
Solving equation \eqref{eq:EE:AIPTW-AIPCW} gives the estimator defined
in equation \eqref{eq:ATE:AIPTW-AIPCW}. Similar derivations for the
\(\IPTW\) estimator in presence of censoring lead to equation
\eqref{eq:ATE:IPTW-AIPCW}.

\subsection{Proof of theorem 1}
\label{appendix:consistency}
\textbf{Correctly specified censoring model}: in this case \(\Cmodel^*\) and
\(M^{c,*}\) are equal to \(\Cmodel\) and \(M^{c}\), respectively. We
use the second notation and show that
\(\widehat{\ATE}_{\AIPTW,\AIPCW}(\tau)\) and
\(\widehat{\ATE}_{\AIPTW}(\tau)\) have the same large sample limit.
We denote by \(\mathcal{F}_{t,i}\) the natural history up to time
\(t\) for individual i where
\(\mathcal{F}_{0,i}=(\Treat_i,\X_i)\). For \(i\in \{1,\ldots,n\}\),
\(M_i^{C}(t)\) is a martingale satisfying \(M_i^{C}(0)=0\). Since
\(\Ymodel^*(\tau|\Treat_i,\X_i)\), \(\Ymodel^*(t|\Treat_i,\X_i)\),
\(\Smodel^*(t|\Treat_i,\X_i)\), and \(\Cmodel(t|\Treat_i,\X_i)\) are
predictable with respect to \(\mathcal{F}_{0,i}\), we obtain that:

\begin{align*}
\bar{I}(\Tobs_i,\tau|\Treat_i,\X_i) = \int_0^{\Tobs_i \wedge \tau} \frac{\Ymodel^*(\tau|\Treat_i,\X_i)-\Ymodel^*(t|\Treat_i,\X_i)}{\Smodel^*(t|\Treat_i,\X_i)} \frac{1}{\Cmodel(t|\Treat_i,\X_i)} dM_{i}^{\C}(t)
\end{align*}
is a martingale. Using that
\(\Esp[M_i^{C}(t)|\mathcal{F}_{0,i}]=0\), we get:
\begin{align*}
\Esp[\bar{I}(\Tobs_i,\tau|\Treat_i,\X_i)] =
\Esp[\Esp[\bar{I}(\Tobs_i,\tau|\Treat_i,\X_i)|\mathcal{F}_{0,i}]]=0.
\end{align*}
Therefore:
\begin{align*}
\Esp\left[ \bar{I}(\Tobs,\tau|\Treat,\X) \left(\frac{\Treat}{\Tmodel^*(\X)} - \frac{1-\Treat}{1-\Tmodel^*(\X)}\right) \right] 
= \Esp\left[ \Esp[ \bar{I}(\Tobs,\tau|\Treat,\X) | \mathcal{F}_0 ] \left(\frac{\Treat}{\Tmodel^*(\X)} - \frac{1-\Treat}{1-\Tmodel^*(\X)}\right) \right] 
= 0
\end{align*}
where the outer expectation is taken over the joint distribution of
\(\Tobs,\Treat\) and \(\X\). Moreover:
\begin{align*}
&\Esp\left[\frac{\Ind[\Tobs \leq \tau, \Etype \neq 0]}{\Cmodel(\Tobs|\Treat,\X)}
\Y(\tau) \left(
\frac{\Treat}{\Tmodel^*(\X)} 
- \frac{1-\Treat}{1-\Tmodel^*(\X)}
\right)\right] \\
&= \Esp\left[\Esp\left[\frac{\Ind[ C > \Tobs \wedge \tau]}{\Cmodel(\Tobs \wedge \tau|\Treat,\X)} \Bigg| \Treat, \X \right]
\Esp\left[ \Y(\tau) \left(
\frac{\Treat}{\Tmodel^*(\X)} 
- \frac{1-\Treat}{1-\Tmodel^*(\X)}
\right) \Bigg| \Treat, \X \right] \right] \\
&= \Esp\left[ \Y(\tau) \left(
\frac{\Treat}{\Tmodel^*(\X)} 
- \frac{1-\Treat}{1-\Tmodel^*(\X)}
\right) \right],
\end{align*}
where we have used the conditional independent censoring
assumption. So \(\lim_{n \rightarrow \infty}
\widehat{\ATE}_{\AIPTW,\AIPCW}(\tau) = \lim_{n \rightarrow \infty}
\widehat{\ATE}_{\AIPTW}(\tau)\) and 1. and 2. follow from the double
robustness of \(\widehat{\ATE}_{\AIPTW}(\tau)\).

\textbf{Misspecified censoring model}: We assume that the outcome model and
survival model are correctly specified, i.e., \(\Ymodel^*=\Ymodel\) and
\(\Smodel^*=\Smodel\). Using equation \eqref{eq:magic}, we obtain:
\begin{align*}
&\frac{\Y_i(\tau)\Ind[\Tobs_i \leq \tau, \Etype \neq 0]}{\Cmodel^*(\Tobs_{i} |\Treat_i,\X_i)} + \int_0^{\Tobs_i \wedge \tau} \frac{\Ymodel^*(\tau|\Treat_i,\X_i)-\Ymodel^*(t|\Treat_i,\X_i)}{\Smodel^*(t|\Treat_i,\X_i)\Cmodel^*(t|\Treat_i,\X_i)} dM_i^{\C,*}(t) \\
=& \Y_i(\tau) + \int_0^{\Tobs_i \wedge \tau} \frac{\frac{\Ymodel(\tau|\Treat_i,\X_i)-\Ymodel(t|\Treat_i,\X_i)}{\Smodel(t|\Treat_i,\X_i)} - \Y_i(\tau)}{\Cmodel^*(t|\Treat_i,\X_i)} dM_i^{\C,*}(t) \\
=& \Y_i(\tau) + \int_0^{\Tobs_i \wedge \tau} \frac{\Esp[\Y_i(\tau)|\T_i>t,\Treat_i,\X_i] - \Y_i(\tau)}{\Cmodel^*(t|\Treat_i,\X_i)} dM_i^{\C,*}(t).
\end{align*}
We now show that the second term has null expectation. Denoting
\(\epsilon_i(t) = \frac{\Esp[\Y_i(\tau)|\T_i>t,\Treat_i,\X_i)] -
\Y_i(\tau)}{\Cmodel^*(t|\Treat_i,\X_i)}\) and using the conditional
independent censoring assumption, we have that
\(\Esp[\epsilon_i(t)|\T_i>t,\Treat_i,\X_i]=0\). 
With \(R_i(t)\) the at risk process, we can decompose the second term further in two terms:
\begin{equation}\label{eq:jegkanikkemere}
\int_0^{\Tobs_i \wedge \tau} \epsilon_i(t) dM_i^{\C,*}(t) = \int_0^{\Tobs_i \wedge \tau} \epsilon_i(t) dM_i^{\C}(t) + 
	\int_0^{\Tobs_i \wedge \tau} \epsilon_i(t) R_i(t) 
d\left(\Lambda^{\C}(t|\Treat_i,\X_i) - \Lambda^{\C,*}(t|\Treat_i,\X_i)\right)
\end{equation}
The first term in \eqref{eq:jegkanikkemere}  is a mean-zero martingale, and the second term in equation \eqref{eq:jegkanikkemere} also has mean zero 
since

\begin{align*}
&  \Esp\left[ \int_0^{\Tobs_i \wedge \tau} R_i(t) \epsilon_i(t) \left(\lambda^{\C}(t|\Treat_i,\X_i) - \lambda^{\C,*}(t|\Treat_i,\X_i)\right) dt \Big| \Treat_i,\X_i  \right] \\
&= \Esp\left[ \int_0^{\tau} R_i(t) \epsilon_i(t) \left(\lambda^{\C}(t|\Treat_i,\X_i) - \lambda^{\C,*}(t|\Treat_i,\X_i)\right) dt \Big| \Treat_i,\X_i  \right] \\
&=  \int_0^{\tau} \Esp\left[R_i(t) \epsilon_i(t)| \Treat_i,\X_i  \right] \left(\lambda^{\C}(t|\Treat_i,\X_i) - \lambda^{\C,*}(t|\Treat_i,\X_i)\right) dt = 0,
\end{align*}
because \(\Esp\left[R_i(t) \epsilon_i(t) | \Treat_i,\X_i \right] =
\Esp\left[R_i(t) \Esp[\epsilon_i(t) | \T_i>t,\Treat_i,\X_i] |\Treat_i,\X_i
\right]=0\).  So the large sample limit of the \(\AIPTW,\AIPCW\)
estimator is:
\begin{align*}
&\lim_{n \rightarrow \infty} \widehat{\ATE}_{\AIPTW,\AIPCW}(\tau) = \Esp\Bigg[
\Ymodel(\tau|\Treat=1,\X) - \Ymodel(\tau|\Treat=0,\X)  \\
&  + \left(
\frac{\Treat}{\Tmodel^*(\X)} 
- \frac{1-\Treat}{1-\Tmodel^*(\X)}
\right) \left( \Y(\tau) - \Ymodel(\tau|\Treat,\X) + \int_0^{\Tobs \wedge \tau} \epsilon(t) dM^{\C,*}(t) \right) \Bigg] \\
&= \Esp[\Ymodel(\tau|\Treat=1,\X) - \Ymodel(\tau|\Treat=0,\X)] \\
& + \Esp\left[\left(
\frac{\Treat}{\Tmodel^*(\X)} 
- \frac{1-\Treat}{1-\Tmodel^*(\X)}
\right) \left( \Esp[\Y(\tau)|\Treat,\X] - \Ymodel(\tau|\Treat,\X) + \Esp\left[\int_0^{\Tobs \wedge \tau} \epsilon(t) dM^{\C,*}(t) \Bigg| \Treat, \X \right] \right) \right] \\
&= \Esp[\Ymodel(\tau|\Treat=1,\X) - \Ymodel(\tau|\Treat=0,\X)]
\end{align*}
which also equals \(\ATE(\tau)\).

\subsection{Influence function of the AIPTW,AIPCW estimator}
\label{appendix:iid}
We define the functional \(\nu\) as a mapping of a set of probability
measures to the real numbers such that for a probability measure
\(\Prob\):
\begin{align*}
\nu(\Prob) = \nu_1(\Ymodel,\Tmodel,\Smodel,\Cmodel,\Prob) - \nu_0(\Ymodel,\Tmodel,\Smodel,\Cmodel,\Prob)
\end{align*}
where, for \(a \in \{0,1\}\) and denoting \(\Tmodel^a(\X) = a \Tmodel(\X) + (1-a) (1-\Tmodel(\X))\), we have:
\begin{align*}
& \nu_a(\Ymodel,\Tmodel,\Smodel,\Cmodel,\Prob) = \Esp\Bigg[ \Ymodel(\tau|\Treat = a, \X)  \\
& + \frac{\Ind[\Treat=a]}{\Tmodel^a(\X)}\left(Y(\tau) - \Ymodel(\tau|\Treat,\X) + \int_0^{\Tobs \wedge \tau} \frac{\frac{\Ymodel(\tau|\Treat,\X) - \Ymodel(t|\Treat,\X)}{\Smodel(t|\Treat,\X)} - \Y(\tau)}{\Cmodel(t|A,W)} dM^{C}(t) \right) \Bigg].
\end{align*}
where the expectation is relative to the joint distribution of
\(\Treat\) and \(\X\). By denoting \(\Prob_n\) the empirical
distribution function we have that
\(\nu(\Prob_n)=\widehat{\ATE}_{\AIPTW,\AIPCW}(\tau)\). So to obtain
the influence function of the \(\AIPTW,\AIPCW\) estimator, we only
need to derive the influence function associated with the estimator of
\(\nu_a\). Using Slutsky theorem, one can show that
\(\nu_a(\hatYmodel,\hatTmodel,\hatSmodel,\hatCmodel,\Prob_n)-\nu_a(\Ymodel^*,\hatTmodel,\hatSmodel,\hatCmodel,\Prob_n)\)
converges towards
\(\nu_a(\hatYmodel,\Tmodel^*,\Smodel^*,\Cmodel^*,\Prob_n)-\nu_a^*\)
where \(\nu_a^* =
\nu_a(\Ymodel^*,\Tmodel^*,\Smodel^*,\Cmodel^*,\Prob)\). Expanding with
respect to each argument leads to:
\begin{align*}
\sqrt{n} &\left(\nu_a(\hatYmodel,\hatTmodel,\hatSmodel,\hatCmodel,\Prob_n)-\nu_a^*\right) = \sqrt{n} \left(\nu_a(\hatYmodel,\Tmodel^*,\Smodel^*,\Cmodel^*,\Prob)-\nu_a^*\right) \\
&\quad + \sqrt{n} \left(\nu_a(\Ymodel^*,\hatTmodel,\Smodel^*,\Cmodel^*,\Prob)-\nu_a^*\right) + \sqrt{n} \left(\nu_a(\Ymodel^*,\Tmodel^*,\hatSmodel,\Cmodel^*,\Prob)-\nu_a^*\right) \\
& \quad + \sqrt{n} \left(\nu_a(\Ymodel^*,\Tmodel^*,\Smodel^*,\hatCmodel,\Prob)-\nu_a^*\right) + \sqrt{n} \left(\nu_a(\Ymodel^*,\Tmodel^*,\Smodel^*,\Cmodel^*,\Prob_n)-\nu_a^*\right)
+ o_p(1).
\end{align*}
We can then calculate the influence function corresponding to each
term: 
\begin{align*}
\sqrt{n} \left(\nu_a(\hatYmodel,\Tmodel^*,\Smodel^*,\Cmodel^*,\Prob)-\nu_a^*\right)
&= \frac{1}{\sqrt{n}} \sum_{i=1}^n \widetilde\IF_{\nu_a,\Ymodel}(\tau;\sample_i) + o_p(1)\\
\sqrt{n} \left(\nu_a(\Ymodel^*,\hatTmodel,\Smodel^*,\Cmodel^*,\Prob)-\nu_a^*\right)
&= \frac{1}{\sqrt{n}} \sum_{i=1}^n \widetilde\IF_{\nu_a,\Tmodel}(\tau;\sample_i)  + o_p(1)\\
\sqrt{n} \left(\nu_a(\Ymodel^*,\Tmodel^*,\hatSmodel,\Cmodel^*,\Prob)-\nu_a^*\right)
&= \frac{1}{\sqrt{n}} \sum_{i=1}^n \widetilde\IF_{\nu_a,\Smodel}(\tau;\sample_i)  + o_p(1)\\
\sqrt{n} \left(\nu_a(\Ymodel^*,\Tmodel^*,\Smodel^*,\hatCmodel,\Prob)-\nu_a^*\right) 
&= \frac{1}{\sqrt{n}} \sum_{i=1}^n \widetilde\IF_{\nu_a,\Cmodel}(\tau;\sample_i)  + o_p(1)\\
\sqrt{n} \left(\nu_a(\Ymodel^*,\Tmodel^*,\Smodel^*,\Cmodel^*,\Prob_n)-\nu_a^*\right)
&= \frac{1}{\sqrt{n}} \sum_{i=1}^n \widetilde\IF_{\nu_a,\Prob}(\tau;\sample_i)  + o_p(1).
\end{align*}
For instance, writing the difference between
\(\nu_a(\Ymodel^*,\Tmodel^*,\Smodel^*,\Cmodel^*,\Prob_n)\) and
\(\nu_a^*\) gives:
\begin{align*}
&\widetilde{\IF}_{\nu_a,\Prob}(\tau;\sample_i) = \Ymodel^*(\tau|\Treat = a, \X_i) -\nu_a^*  \\
& + \frac{\Ind[\Treat_i=a]}{\Tmodel^{a,*}(\X_i)}\left(Y_i(\tau) - \Ymodel^*(\tau|\Treat_i,\X_i) + \int_0^{\Tobs_i \wedge \tau} \frac{\frac{\Ymodel^*(\tau|\Treat_i,\X_i) - \Ymodel^*(t|\Treat_i,\X_i)}{\Smodel^*(t|\Treat_i,\X_i)} - \Y_i(\tau)}{\Cmodel^*(t|A_i,W_i)} dM_i^{C,*}(t) \right).
\end{align*}
For \(\widetilde\IF_{\nu_a,\Ymodel}\) we use that:
\begin{align*}
& \sqrt{n} \left(\nu_a(\hatYmodel,\Tmodel^*,\Smodel^*,\Cmodel^*,\Prob)-\nu^*_a\right) =\\
& \frac{1}{\sqrt{n}} \sum_{i=1}^n \Esp\Bigg[ \IF_{\Ymodel^*}(\tau,a,\X;\sample_i) \left(1 - \frac{\Ind[\Treat=a]}{\Tmodel^{a,*}(\X)}\right)  \\
& \quad + \frac{\Ind[\Treat=a]}{\Tmodel^{a,*}(\X)} \int_0^{\Tobs \wedge \tau} 
\frac{\IF_{\Ymodel^*}(\tau,\Treat,\X;\sample_i) - \IF_{\Ymodel^*}(t,\Treat,\X;\sample_i) }{\Smodel^*(t|\Treat,\X) \Cmodel^*(t|\Treat,\X)} dM^{\C,*}(t) \Bigg| \sample_i \Bigg]
\end{align*}
where \(\Tmodel^{a,*}\) indicates the large sample limit of
\(\Tmodel^{a}\) and \(\IF_{\Ymodel^*}(t,\Treat,\X,\sample_i)\) is defined in equation \eqref{eq:vonmises}.
This leads to:
\begin{align*}
\widetilde\IF_{\nu_a,\Ymodel}(\tau;\sample_i) =& \Esp\Bigg[ \IF_{\Ymodel^*}(\tau,a,\X;\sample_i) \left(1 - \frac{\Ind[\Treat=a]}{\Tmodel^{a,*}(\X)}\right) \\
& \qquad + \frac{\Ind[\Treat=a]}{\Tmodel^{a,*}(\X)} \int_0^{\Tobs \wedge \tau} 
\frac{\IF_{\Ymodel^*}(\tau,\Treat,\X;\sample_i) - \IF_{\Ymodel^*}(t,\Treat,\X;\sample_i) }{\Smodel^*(t|\Treat,\X) \Cmodel^*(t|\Treat,\X)} dM^{\C,*}(t) \Bigg| \sample_i \Bigg].
\end{align*}
Similar derivations give:
\begin{align*}
&\widetilde{\IF}_{\nu_a,\Tmodel}(\tau;\sample_i) = -\Esp\Bigg[ \IF_{\Tmodel^*}(\X;\sample_i)\frac{\Ind[\Treat=a]}{\left(\Tmodel^{a,*}(\X)\right)^2} \Bigg(\frac{\Y(\tau)\Ind[\Tobs \leq \tau|\Etype \neq 0]}{\Cmodel^*(\Tobs|\Treat,\X)} - \Ymodel^*(\tau|\Treat,\X) \\
& \qquad \qquad \qquad \qquad  \qquad \qquad \qquad \qquad \qquad + \int_0^{\Tobs \wedge \tau}  \frac{\Ymodel^*(\tau|\Treat,\X) - \Ymodel^*(t|\Treat,\X)}{\Smodel^*(t|\Treat,\X) \Cmodel^*(t|\Treat,\X)} dM_i^{\C,*}(t) \Bigg) \Bigg| \sample_i \Bigg] \\
&\widetilde{\IF}_{\nu_a,\Smodel}(\tau;\sample_i) = -\Esp\Bigg[ \frac{\Ind[\Treat=a]}{\Tmodel^{a,*}(\X)} \int_0^{\Tobs \wedge \tau} 
\IF_{\Smodel^*}(t,\Treat,\X;\sample_i) \frac{\Ymodel^*(\tau|\Treat,\X) - \Ymodel^*(t|\Treat,\X)}{\Smodel^*(t|\Treat,\X)^2 \Cmodel^*(t|\Treat,\X)} dM^{\C,*}(t) \Bigg| \sample_i \Bigg] \\
&\widetilde{\IF}_{\nu_a,\Cmodel}(\tau;\sample_i) = -\Esp\Bigg[ \frac{\Ind[\Treat=a]}{\Tmodel^{a,*}(\X)} \left(
 \IF_{\Cmodel^*}(\Tobs,\Treat,\X;\sample_i) \frac{\Y(\tau)\Ind[\Tobs \leq \tau|\Etype \neq 0]}{\Cmodel^*(\Tobs|\Treat,\X)^2} \right. \\
& \qquad \qquad \qquad \qquad \qquad \qquad+ \int_0^{\Tobs \wedge \tau} \IF_{\Cmodel^*}(t,\Treat,\X;\sample_i) \frac{\Ymodel^*(\tau|\Treat,\X) - \Ymodel^*(t|\Treat,\X)}{\Smodel^*(t|\Treat,\X)\Cmodel^*(t|\Treat,\X)^2} dM^{\C,*}(t) \\
& \qquad \qquad \qquad \qquad \qquad \qquad \left. + \int_0^{\Tobs \wedge \tau} 
\frac{\Ymodel^*(\tau|\Treat,\X) - \Ymodel^*(t|\Treat,\X)}{\Smodel^*(t|\Treat,\X) \Cmodel^*(t|\Treat,\X)} d\left(\IF_{\Lambda^{C,*}}(t,\Treat,\X;\sample_i)\right) \right) \Bigg| \sample_i \Bigg] \\
\end{align*}
where \(\IF_{\Lambda^{C,*}}\) denotes the influence function of the
cumulative hazard associated to the censoring mechanism. Furthermore, denoting 
\begin{align*}
\IF_{\AIPTW,\AIPCW}(\tau; \sample_i) =
\widetilde\IF_{\AIPTW,\AIPCW}(\tau; \sample_i)
+ \phi_{\AIPTW,\AIPCW}(\tau;\sample_i;\IFY,\IFT,\IFS,\IFC)
\end{align*}
the influence function of the \(\AIPTW,\AIPCW\) estimator, we get that
\(\phi_{\AIPTW,\AIPCW}(\tau;\sample_i;\IFY,\IFT,\IFS,\IFC)\) equals
\begin{align*}
& \widetilde\IF_{\nu_1,\Ymodel}(\tau;\sample_i) + \widetilde\IF_{\nu_1,\Tmodel}(\tau;\sample_i) + \widetilde\IF_{\nu_1,\Smodel}(\tau;\sample_i) + \widetilde\IF_{\nu_1,\Cmodel}(\tau;\sample_i)  \\
& - \widetilde\IF_{\nu_0,\Ymodel}(\tau;\sample_i) - \widetilde\IF_{\nu_0,\Tmodel}(\tau;\sample_i) - \widetilde\IF_{\nu_0,\Smodel}(\tau;\sample_i) - \widetilde\IF_{\nu_0,\Cmodel}(\tau;\sample_i) 
\end{align*}
and \(\widetilde\IF_{\AIPTW,\AIPCW}(\tau;\sample_i) = \widetilde\IF_{\nu_1,\Prob}(\tau;\sample_i) - \widetilde\IF_{\nu_0,\Prob}(\tau;\sample_i)\).

Following the same reasoning as in the section \ref{appendix:consistency} and using the
conditional independence between the censoring mechanism and the
treatment variable, we note that:
\begin{itemize}
\item \(\widetilde{\IF}_{\nu_a,\Ymodel}=0\) when the treatment and censoring models are correctly specified.
\item \(\widetilde{\IF}_{\nu_a,\Tmodel}=0\) when the outcome and the censoring models are correctly specified.
\item \(\widetilde{\IF}_{\nu_a,\Smodel}=0\) when the censoring is correctly specified.
\item \(\widetilde{\IF}_{\nu_a,\Cmodel}=0\) when the outcome, survival, and censoring models are correctly specified.
\end{itemize}
So when all models are correctly specified
\(\phi_{\AIPTW,\AIPCW}(\tau;\sample_i,\IF_{\Ymodel^*},\IF_{\Tmodel^*},\IF_{\Smodel^*},\IF_{\Cmodel^*})=0\)
and \(\IF_{\AIPTW,\AIPCW}(\tau;\sample_i) = \widetilde{\IF}_{\AIPTW,\AIPCW}(\tau;\sample_i)\).
\end{document}